\documentclass[preprint,10pt]{elsarticle}

\usepackage{graphicx}
\usepackage{amssymb}

\usepackage{lineno}

\journal{arxiv}

\begin{document}

\begin{frontmatter}


\title{Hyperbolic Center of Mass for a System of Particles on a two-dimensional Space with Constant Negative Curvature: An Application to the Curved $2$- and $3$-Body Problems}



\author{Pedro P. Ortega Palencia}
\address{Universidad de Cartagena, portegap@unicartagena.edu.co}

\author{J. Guadalupe Reyes Victoria}
\address{Universidad Aut\'onoma Metropolitana, revg@xanum.uam.mx}

\begin{abstract}
In this article is given a simple expression for the \textit{ center of mass} for a system of material points in a two-dimensional surface of  constant negative Gaussian curvature. Using basic techniques of Geometry, an expression in intrinsic coordinates  is obtained, and it is showed how it extends the definition for the Euclidean case. The argument is constructive and also serves for defining center of mass of a system of particles on the one-dimensional hyperbolic space  $\mathbb{L}^1_R$. Finally, is showed some applications to the curved $2$- and $3$-body problems. 
\end{abstract}


\end{frontmatter}


\section{Introduction}
\label{S:1}

Center of mass (center of gravity or centroid) is a fundamental concept, and its geometrical and mechanics properties are very important for the understanding of many physical phenomena. Its definition for Euclidean spaces is elemental, nevertheless a definition  for curved spaces is few frequent. In \cite{Galperin}, the author makes an extensive explanation showing the possibility of it concept can be correctly defined in more general spaces, and he signalize the difficulties to defined in spaces of non zero curvature, by lacking of linear structure of ones. While it is true that the author synthesizes the basic properties of the center of mass, in his approach appear some ones without physical meaning, such as the non conservation of total mass of system or the presence of infinities velocities under normal conditions . In \cite{Naranjo} appear a definition of center of mass for two particles in a hyperbolic space, in the same direction to the one presented here, but the authors do not give an expression to calculate it. In \cite{Diacu1} the author makes mention about of the  difficulty for defining of center of mass in curved spaces. He provides a class of orbits in the curved $n$-body problem for which, \textquotedblleft no point that could play the role of the center of mass is fixed or moves uniformly along a geodesic''. This proves that the equations of motion lack center-of-mass and linear-momentum integrals. But nevertheless, he is not provide a way to calculate or determinate this element.

In this article, the problem of give a mathematical expression  for computing the center of mass of a system of $n$ particles sited on the upper sheet of two-dimensional hyperbolic sphere $\mathbb{L}^2_R=\{(x,y,z):\; x^{2}+y^{2}-z^{2}= -R^{2}, z>0\}$ is considered. The formula obtained here, extend of natural way the expression for Euclidean case, contrary to the opinion expressed in \cite{Diacu3}. Through stereographic projection of upper sheet of $\mathbb{L}^2_R$ on the Poincar\'e disk $\mathbb{D}_R^2= \{(x,y):\; x^{2}+y^{2} < R^{2}\}=\{w\in \mathbb{C}:|w|<R\}$, endowed with the conformal metric (see \cite{Diacu2}). 
\begin{equation}\label{met-c}
   ds^2= \frac{4R^4 \,dw d\bar{w}}{(R^2 - |w|^2)^2},
\end{equation}

$\mathbb{D}_R^2$ with the metric (\ref{met-c}) and $\mathbb{L}^2_R$ with the metric of Minkowski space, have the same Gaussian curvature $K=-1/R^2$, and from the Minding's Theorem, both surfaces belong to the isometric differentiable class (see\cite{Dub}).
Following the basic methods of the geometry, we obtain here the expression for the center of mass for a system of $n$ particles sited in the hyperbolic sphere $\mathbb{L}^2_R$ with arbitrary$R$. 

This article is organized as follow: in section 1, are introduced some concepts relative to center of mass in the euclidean spaces. In section 2, are remembered some properties of stereographical projection and it is proceeded to deduce the expression for the center of mass, for two particles on the upper branch of hyperbola , from the \textquotedblleft hyperbolic rule of the lever"(see \cite{Galperin}) extended to $\mathbb{L}^2_R$. Once obtained the expression for the center of mass for two particles in $\mathbb{L}^1_R$, it can be extended naturally to a system of $n$ particles in $\mathbb{L}^1_R$, and the same way, to a system of $n$ particles in $\mathbb{L}^2_R$. The expression obtained here, satisfies the five axioms for the \textquotedblleft Axiomatic Centroid" established in \cite{Galperin}.

\section{ One-dimensional Euclidean case}

Let consider two particles with positive masses sited in the real line at the points $x_{1}$ and $x_{2}$. The (Euclidean) center of mass of system is defined be the point $x_{c}$ 

\begin{equation}
x_{c}=\frac{m_{1}x_{1}+m_{2}x_{2}}{m_{1}+m_{2}}
\end{equation}
A direct computing shows that $m_{1}|x_{c}-x_{1}|=m_{2}|x_{c}-x_{2}|$(Euclidean rule of the lever). It is easy to prove that $x_c$ is the unique point in the segment (geodesic) joining $x_{1}$ and $x_{2}$ with this property. This definition can be extend to more dimensions, in Euclidean spaces. But nevertheless, this definition not can be extended to spaces in general, because is possible that in such spaces is not defined a linear structure. But with the \textquotedblleft rule of the lever" in mind is possible carries this definition to Riemannian  surfaces, as we shall see later.

\section{Center of mass in a two-dimensional hyperbolic space}

\subsection{Some observations about the Stereographic Projection of Hyperbolic Sphere on the Poincar\'e disk}

Let $P:\mathbb{L}^2_R \to \mathbb{D}_R^2$ the stereographic projection, then for $(x,y,z) \in \mathbb{L}^2_R$, we have $P(x,y,z)= w=u+iv$ where $u= \frac{Rx}{R+z} $ and $v=\frac{Ry}{R+z}$ and moreover the inverse projection is $P^{-1}:\mathbb{D}_R^2 \to \mathbb{L}^2_R$ and (see \cite{Diacu2}),

$$ P^{-1}(u+iv)=\left(\frac{2R^2u}{R^2-u^2-v^2},\frac{2R^2v}{R^2-u^2-v^2}, \frac{R(u^2+v^2-R^2)}{R^2-u^2-v^2}\right)$$

 the point $(0,0,R)$ will called \textit{south pole}. $P^{-1}$ transform lines through the origin in \textit{meridians} (hyperbolas through the south pole and parallels to $z$ axis) and the circles with center in origin, $\{w\in \mathbb{D}_R^2: |w|= const.<R\}$ in \textit{parallels} (concentric circles on the upper sheet of hyperboloid). 

If we consider the stereographic projection of the one-dimensional hyperbolic sphere $\mathbb{L}^1_R$ on the one-dimensional Poincar\'e disk $\mathbb{D}_R^1=(-R,R)$ with the induced metric from (\ref{met-c}), then the above equations are reduced to:

$P(x,y)=u$ where $u= \frac{Rx}{R+y} $ and $u\in (-R,R)$ and moreover the inverse projection is, 

$$ P^{-1}(u)=\left(\frac{2R^2u}{R^2-u^2},\frac{R(R^2+u^2)}{R^2-u^2}\right)$$

In this last case, the length of arc from the south pole $P_s$ to arbitrary point $(x,y)$ is:

$$s= \int_0^u \frac{2R^2dt}{R^2-t^2}=R\ln \left(\frac{R+u}{R-u}\right)$$

More general, the length of arc $s$ from the point $Q_1(x_1,y_1)$ to $Q_2(x_2,y_2)$ in the same parallel, if their stereographical projections are $u_1$ and $u_2$, is

$$s=R\left(\ln \left(\frac{R+u_2}{R-u_2}\right)-\ln \left(\frac{R+u_1}{R-u_1}\right)\right)$$

Consider now two masses $m_1,m_2$ sited in the points $Q_1,Q_2$ respectively, and  

let $Q_{c}(x_{c},y_{c})$ the coordinates of hyperbolic center of mass, and $s_1$ the length of arc from $Q_1$ to $Q_{c}$ and $s_2$ the length of arc from $Q_{c}$ to $Q_2$ (see Fig.1), then, from the relation (hyperbolic rule of the lever) $m_1s_1=m_2s_2$ it is follows:

$$Rm_1\left(\ln \left(\frac{R+u_{c}}{R-u_c}\right)-\ln \left(\frac{R+u_1}{R-u_1}\right)\right)=Rm_2\left(\ln \left(\frac{R+u_2}{R-u_2}\right)-\ln \left(\frac{R+u_{c}}{R-u_c}\right)\right)$$

\begin{figure}[h]
\centering\includegraphics[width=0.6\linewidth]{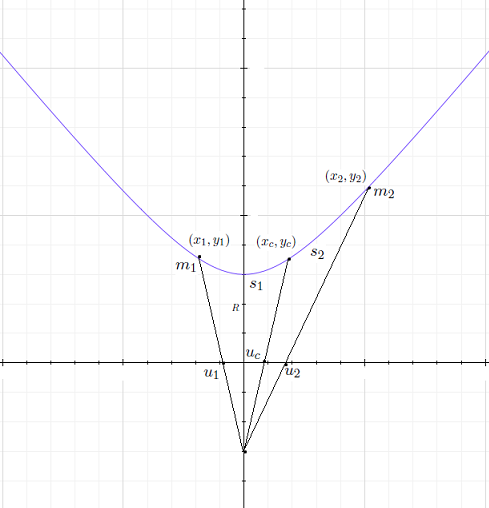}
\caption{Center of Mass on the one-dimensional Hyperbolic Sphere}
\end{figure}

Therefore,

\begin{equation}
\ln \left(\frac{R+u_{c}}{R-u_c}\right)=\frac{1}{m_1+m_2}\left(m_1\ln \left( \frac{R+u_1}{R-u_1}\right)+m_2\ln \left( \frac{R+u_2}{R-u_2}\right)\right)
\end{equation} 
 This concept can be extended a $n$ particles $(x_1,y_1)$,$(x_2,y_2)$,...,$(x_n,y_n)$ with masses $m_1$,$m_2$,...,$m_n$ sited on $\mathbb{L}^1_R$ and with stereographical projections $u_1$,$u_2$,...,$u_n$ in the one-dimensional Poincar\'e disk $(-R,R)$ in the following way:
 
\begin{equation}
\label{cme}
\ln \left(\frac{R+u_c}{R-u_c}\right)=\frac{1}{m}\sum_{k=1}^n m_k\ln \left(\frac{R+u_k}{R-u_k}\right)
\end{equation}
where, $m=\sum_{k=1}^n m_k$

\subsection{Center of mass for a system of $n$ particles in $\mathbb{L}^2_R$}
Now, we extend the \textquotedblleft rule of the lever" to context more general:  let a Riemannian surface $T$ and two particles with masses $m_1,m_2$ sited in the points $\tau_1,\tau_2\in T$, respectively, then the $T$-center of mass is defined the point $\tau_c$ in the geodesic joining $\tau_1$ to $\tau_2$ such that is verified the following relation: $$m_1d(\tau_1,\tau_c)=m_2d(\tau_2,\tau_c)$$ where $d$ is the metric in $T$. For the case of $\mathbb{L}^2_R$, geodesics are hyperbolas determined for the intersection of upper sheet of hyperboloid with the plane draw for the pair of points and the origin $(0,0,0)$.\\

Inductively we can extend in natural way the concept of center of mass to a system of $n$ particles (see \cite{Galperin}, Axiom 2 (induction axiom)) , and using a similar argument bellow, to obtain the next:

\textbf {Proposition 1}: \textit{ Let $m_1$,$m_2$,....,$m_n$, $n$ masses particles sited respectively in the points $(x_1,y_1,z_1)$,$(x_2,y_2,z_2)$,..., $(x_n,y_n,z_n)$ on  $\mathbb{L}^2_R$ with stereographical projections  $w_1$,$w_2$,..., $w_n$, in the Poincar\'e disk $\mathbb{D}^2_R$ , and let $w_c \in \mathbb{D}^2_R$  the stereographical proyection of the hyperbolic center of mass $(x_c,y_c,z_c) \in \mathbb{L}^2_R$  , then it is satisfy the following equation:}

\begin{equation}
\label{cme}
\ln \left(\frac{R+w_c}{R-w_c}\right)=\frac{1}{m}\sum_{k=1}^n m_k\ln \left(\frac{R+w_k}{R-w_k}\right)
\end{equation}
where $m=\sum_{k=1}^n m_k$

or equivalently

\begin{equation}
\label{cmh}
 \left(\frac{R+w_c}{R-w_c}\right)^m=\prod_{k=1}^n \left(\frac{R+w_k}{R-w_k}\right)^{m_k}
\end{equation}

If in each fraction is divided their numerator and denominator for $R$ and both side rises to the power $R$, when  $R\rightarrow \infty$, it is obtained

$$\exp \left( 2mw_c \right)= \exp \left(2\sum_{k=1}^n m_k w_k \right)$$

or equivalently,

\begin{equation}
\label{cmee}
w_c=\frac{1}{m}\sum_{k=1}^n m_k w_k
\end{equation}
And this corresponds to the equation for the center of mass in the Euclidean complex plane, that is, the complex plane (or $\mathbb{R}^2$), with Euclidean metric and zero curvature. So, we have obtained the following:

\textbf{Corollary} : \textit{The expressions (5) or (6) extend the concept of Euclidean center of mass , to hyperbolic spaces with constant curvature.} 

\section{Some applications to the curved $2$ and $3$-body problems}

In \cite{Diacu2} is studied the curved $n$-body problem in a two-dimensional space with negative constant curvature, and is considered the model of $\mathbb{L}^2_R$. In particular, for the configurations called \textit{relative equilibria}, for the $2$ body problem is established the next result:\

\textbf{Theorem 3}. \textit{ Consider 2 point particles of masses $m_1,m_2 > 0$ moving on the Poincar\'e disk $\mathbb{D}^2_R$, whose center is
the origin, $0$, of the coordinate system. Then $z = (z_1, z_2)$ is an elliptic relative equilibrium of system (21) with
$n = 2$, if and only if for every circle centered at $0$ of radius $\alpha$, with $0<\alpha<R$ , along which $m_1$ moves, there is
a unique circle centered at $0$ of radius $r$, which satisfies $0< r < R$ and (46), along which $m_2$ moves, such that,
at every time instant, $m_1$ and $m_2$ are on some diameter of $\mathbb{D}^2_R$, with $0$ between them. Moreover,}
\begin{itemize}
\item[1)] if $m_2 > m_1 > 0$ and $\alpha$ are given, then $ r<\alpha$;
\item[2)] if $m_1 = m_2 > 0$ and ) are given, then $r =\alpha $;
\item[3)] if $m_1 > m_2 > 0$ and ) are given, then $r > \alpha$.
\end{itemize}

This result can be reformulated in a more precise form, using the expression for the hyperbolic center of mass, taking into account that in a configuration correspond to a relative equilibrium it is invariant with the time, because the distance and angles between particles not change, and for this reason is sufficient consider the initial configuration on the $x$ axis (same diameter of disk), and moreover $\alpha$  correspond to the length measure over the Poincar\'e disk, of the projection of arc $s_1$ over the hyperbolic sphere $\mathbb{L}^2_R$, and the $r$ is the length of the projection in the disk one, of the arc $s_2$ over the hyperbolic sphere, then we have the next relations:  

$$s_1= ln \left(\frac{R+\alpha}{R-\alpha}\right)$$ and 
$$s_2= -ln \left(\frac{R+r}{R-r}\right)$$

Substituting in the hyperbolic liver rule $m_1s_1=m_2s_2$ and the expression for the center of mass, we obtain:

$$ \left(\frac{R+\alpha}{R-\alpha}\right)^{m_1}= \left(\frac{R+r}{R-r}\right)^{-m_2} $$

Thus it is follow from equation (6) that

$$ \frac{R-w_c}{R+w_c}= 1$$ it is follows intermediately that $w_c=0$ and so, the center of mass is fixed for every time in the $South Pole$ of the hyperbolic  sphere  $(0,0,R)$ and moreover, Theorem 3 can be expressed in the following form:\\

\textbf{Theorem 3\'}: \textit{For every configuration of elliptic relative equilibrium for the $2$-body problem with masses $m_1,m_2$ sited in the points $P_1(x_1,y_1,z_1)$ and $P_2(x_2,y_2,z_2)$ on the hyperbolic sphere of radius $R$, if $s_1$ and $s_2$ are the length of arcs measure from the South Pole $(0,0,R)$ to the points $P_1$ and $P_2$ respectively, then it is satisfies the relation $m_1s_1=m_2s_2$, and moreover the center of mass of the system is fixed in $(0,0,R)$ for every time.} 

A similar analysis shows, that for the configurations of the hyperbolic $3$-body problem called \textit{Eulerian solutions} and \textit{ Lagrangian solutions}, and characterized by Theorem 4, Proposition 2 and Theorem 5 in \citep{Diacu2} the center of mass is fixed in the point $(0,0,R)$ of $\mathbb{L}_R^2$. Finally, for the configurations called \textit{hyperbolic relative equilibria}  in \cite{Diacu4} , using an argument of symmetries, the center of mass is moving over the hyperbola of intersection of $\mathbb{L}_R^2$ (in such work is used the notation $\mathbb{H}^2$ instead $\mathbb{L}_R^2$)  with the plane $x=0$, and it is a geodesic curve over the surface.

In conclusion, although the hyperbolic center of mass is not a first integral for the curved $n$-body problem, for some types of configurations (relative equilibria, for example) it haves dynamical properties, analogues to the center of mass in the Euclidean case, under the action of Newtonian gravitational potential, this is, it is fixed or moving over a geodesic curve.     

\section{Funding Sources}

This work was supported by the University of Cartagena, Cartagena, Colombia [Grant Number 066-2013] and the Universidad Aut\'onoma Metropolitana, Iztapalapa, Ciudad de M\'exico, M\'exico. 

\section{Acknowledgment}

The first author wishes to thank to Professor Joaqu\'{\i}n Luna Torres, for his valuable suggestions.





\section{References}

\bibliographystyle{model1-num-names}

\bibliography{sample.bib}







\end{document}